# Multi-echo Reconstruction from Partial K-space Scans via Adaptively Learnt Basis

Jyoti Maggu, Prerna Singh and Angshul Majumdar

Indraprastha Institute of Information Technology, Delhi

{jyotim, prernas and angshul}@iiitd.ac.in

Abstract – In multi-echo imaging, multiple T1/T2 weighted images of the same cross section is acquired. Acquiring multiple scans is time consuming. In order to accelerate, compressed sensing based techniques have been proposed. In recent times, it has been observed in several areas of traditional compressed sensing, that instead of using fixed basis (wavelet, DCT etc.), considerably better results can be achieved by learning the basis adaptively from the data. Motivated by these studies, we propose to employ such adaptive learning techniques to improve reconstruction of multi-echo scans. This work will be based on two basis learning models – synthesis (better known as dictionary learning) and analysis (known as transform learning). We modify these basic methods by incorporating structure of the multi-echo scans. Our work shows that we can indeed significantly improve multi-echo imaging over compressed sensing based techniques and other unstructured adaptive sparse recovery methods.

1.  Introduction

Application of compressed sensing (CS) techniques in magnetic resonance imaging is a well known area. It has been developing for the past decade since the publication of seminal work by Lustig et al [1]. Basically the objective is to reduce the data acquisition time. This is achieved by partially sampling the K-space. Mathematically this is expressed as,

$$y = RFx + \eta, \ \eta \sim \ ) \qquad (1)$$

Here $x$ is the underlying image to be reconstructed, $F$ is the Fourier mapping between the image and the K-space, $R$ is the restriction operator which corresponds to the partial sampling process, $y$ is the acquired K-space data and $\eta$ is the system noise known to be Normally distributed.

This is a typical under-determined linear inverse problem since the number of K-space scans acquired are fewer than the size of the image. Hence CS plays a role in recovering $x$. CS assumes that the image is sparse in some fixed basis like wavelet or DCT. Usually the basis is assumed to be orthogonal[1] or tight-framed[2]. In either case the synthesis and analysis equations hold.

$Analysis: \alpha = Sx$ (2a)

$Synthesis: x = S^T \alpha$ (2b)

Hence it is possible to express (1) via the synthesis equation (2b) in the following form,

$y = RFS^T \alpha + \eta$ (3)

Since the transform coefficients are supposed to be sparse, one can use CS to recover them. Usually it is posed as an $l_1$-minimization problem.

$$\min_{\alpha} \|y - RFS^T \alpha\|_2^2 + \lambda \|\alpha\|_1$$ (4)

Once the sparse transform coefficients are obtained, the image is recovered by applying the synthesis equation (2b).

The form (4) is called the synthesis prior formulation. This is because it uses the synthesis equation (2b) to formulate the problem. This form is restrictive, since only orthogonal and tight-frame transforms are amenable to it. This precludes various important sparsifying basis like Gabor or finite difference (total variation). As a linear transform, these basis have an analysis operation but not a synthesis operation. To accommodate these basis one needs to formulate recovery as a co-sparse analysis prior formulation. Instead of solving for the sparse coefficients, the analysis formulation directly solves for the image. The analysis formulation is expressed as,

$$\min_{x} \|y - RFx\|_2^2 + \lambda \|Sx\|_1$$ (5)

In several studies, such as [2], it has been observed that the analysis prior yields better results by able to accommodate more powerful albeit non-orthogonal / tight-frame linear sparsifying operators.

So far we have discussed about recovering single-echo MRI images. Usually for quantitative imaging, multiple T1/T2 weighted images are acquired. From these images, the T1/T2 maps are obtained by curve-fitting techniques. For such scans, as multiple echoes need to be acquired the scan time becomes significantly higher. Therefore a concerted effort need to be made to bring down the scan time. Model based CS techniques have been successful in the past to address this problem. We will briefly discuss these techniques to understand the state-of-the-art in this topic.

The partial K-space scan for each echo can be expressed as follows,

$y_i = R_i F x_i + \eta$, i=1...C (6)

Here $i$ is the echo number and $C$ is the total number of acquired echoes (usually 16 or 32). It is possible to change the sampling mask $R$ for each scan, hence the mask also is indexed by $i$. In a multiple measurement vector form (6) can be expressed as,

---

[1] $Orthogonal: S^T S = I = SS^T$

[2] $Tight - frame: S^T S = I \neq SS^T$

$$vec(Y) = R(I \otimes F)vec(X) + \eta \tag{7}$$

where $Y = [y_1 | ... | y_C]$, $X = [x_1 | ... | x_C]$ and $R$ is formed by $R_i$'s as block diagonals.

It has been argued in prior studies [3-6] that since the underlying structure remains constant in all the images, it is beneficial to exploit this common structure in the CS recovery. This leads to one realization of model based compressed sensing [7].

In [3-6] it has been argued that since wavelet (or any other sparsifying transform) encodes the edge information (high value along edges and near about zeroes elsewhere), as long as the underlying structure remains constant, the transform coefficients in all the echoes will have a common sparse support, i.e. the positions of the non-zero coefficients will remain the same (although their values will vary). In such a scenario, the sparsified transform coefficients of $X$ would be a row-sparse matrix; only those rows corresponding to the edges of the underlying structure would have non-zero values. Under this assumption, [4, 5] reconstructed all the multi-echo images jointly by solving the following optimization problem.

$$\min_A \|vec(Y) - R(I \otimes F)vec(S^T A)\|_2^2 + \lambda \|A\|_{2,1} \tag{8}$$

where $A = SX$ is the matrix of transform coefficients and the $l_{2,1}$-norm is defined as the sum over the $l_2$-norms of the rows. It is the standard prior from signal processing that recovers row-sparse solutions [8-10].

As before (8) is a synthesis formulation. Better results were obtained in [3, 6] by formulating joint recovery of the multi-echo images as an analysis formulation.

$$\min_X \|vec(Y) - R(I \otimes F)vec(X)\|_2^2 + \lambda \|SX\|_{2,1} \tag{9}$$

This is the state-of-the-art in multi-echo reconstruction. The idea of joint recovery has been utilized in other MRI problems in other contexts. For example in [11] it has been used for recovering diffusion weighted images. In [12], similar ideas were exploited for recovering sensitivity encoded images in parallel MRI.

The studies discussed so far assumed that the sparsifying basis is known. Such mathematically defined basis – wavelet, DCT, finite difference etc. can sparsely represent a large class of signals. For examples DCT can be used for compressing, image, videos and speech; so does wavelet. However, it is known that for representing a particular signal the best basis is the one learnt from the signal itself; the basis needs to adapt to the signal. Therefore it is expected that, instead of employing fixed basis, if we learn the basis adaptively from the signal a better recovery will be possible. This constitutes the fundamental idea behind dictionary learning. It learns a dictionary from the signal itself while reconstructing it. In the context of single echo MRI reconstruction [13, 14] it is formulated as,

$$\min_{x,D,Z} \|y - RFx\|_2^2 + \mu \left( \sum_j \|P_j x - Dz_j\|_F^2 + \lambda \|z_j\|_1 \right) \tag{10}$$

As before, $x$ is the image to be recovered. $P_j$ is the patch extraction operator. $D$ is the adaptively learnt sparsifying basis and $z_i$ the coefficients corresponding to the $j^{th}$ patch; $Z$ is formed by stacking all the $z_i$'s as columns. The reconstruction proceeds in two steps (in every iteration). In one step the image is estimated –

$$\min_x \|y - RFx\|_2^2 + \mu \left( \sum_j \|P_j x - Dz_j\|_F^2 \right) \tag{11}$$

This is a simple least squares problem. The next step corresponds to dictionary learning, where the basis and the coefficients are updated.

$$\min_{D,Z} \sum_i \|P_j x - D z_j\|_F^2 + \lambda \|z_j\|_1 \tag{12}$$

It has been shown in [13, 14] that significantly better acceleration factors can be achieved (given the quality) with dictionary learning based reconstruction compared to standard CS. Dictionary learning has also been used for dynamic MRI reconstruction [15, 16].

Dictionary learning is a synthesis formulation, i.e. it learns a dictionary so as to synthesize / generate the original signal from the coefficients. There is an analysis formulation to it called transform learning. In transform learning an operator is learnt such that it analyzes the data to generate the coefficients. The transform learning based formulation for recovering MR images [17, 18] is expressed as follows.

$$\min_{x,T,Z} \|y - RFx\|_2^2 + \mu \left( \sum_j \|T P_j x - z_j\|_F^2 + \lambda \|z_j\|_1 + \gamma \left( \|T\|_F^2 - \log\det T \right) \right) \tag{13}$$

Here $T$ is the analysis transform; the rest of the symbols carry their usual meaning. Note that transform learning is more involved than dictionary learning; this owes to the extra penalty term $-\gamma \left( \|T\|_F^2 - \log\det T \right)$. Note that without this term, transform learning is likely to converge to the trivial solution $T=0$, $Z=0$. The extra penalty term ensures that the trivial solution is prevented by the $-\log\det T$ penalty; the penalty $\|T\|_F^2$ is to balance scale.

As in dictionary learning, the transform learning based solution also proceeds in two stages – image estimation (14) and transform learning (15).

$$\min_x \|y - RFx\|_2^2 + \mu \left( \sum_j \|T P_j x - z_i\|_F^2 \right) \tag{14}$$

$$\min_{T,Z} \sum_j \|T P_j x - z_j\|_F^2 + \lambda \|z_j\|_1 + \gamma \left( \|T\|_F^2 - \log\det T \right) \tag{15}$$

It should be noted that the analysis formulation presented in studies like [19] is not similar to transform learning; it borrows from the analysis dictionary learning formulation [20].

The dictionary learning and transform learning formulations discussed here in can substitute standard CS techniques for single echo MRI reconstruction. However, as discussed before, the multi-echo MRI reconstruction problem needs to exploit structure of the underlying images. Therefore the standard adaptive basis learning solutions will not be very useful in this scenario. The goal of this work is to modify dictionary learning and transform learning formulation into the fold of model based compressed sensing so as to recovery joint sparse images.

A brief review of the state-of-the-art has already been given in the introduction. We describe our proposed techniques in the following section. The experimental results will be given in section 3. The conclusions of this work and future direction of research is discussed in section 4.

## 2. Proposed Approach

We will propose two solutions for solving the multi-echo reconstruction problem. The first one will be based on dictionary learning and the second one based on transform learning.

### 2.1. Dictionary Learning

Our objective is to recover multi-echo T1/T2 weighted images. These images correspond to the same cross section and hence are structurally similar. The T1/T2 weightings reflect as difference in contrast between the tissues. There are two ways, one can attempt to recover these images. The first one is a piecemeal approach, i.e. each of the echoes are separately recovered using dictionary learning (10). However this is not optimal; it does not exploit the common structure across all the echoes.

The other approach is to express the problem in a Kronecker compressed sensing (KCS) framework [21]. In the dictionary learning based formulation it has been used by [22] for denoising color images. In such a formulation, one will express the dictionary in the following fashion,

$$\begin{bmatrix} P_i x_1 \\ \ldots \\ P_i x_C \end{bmatrix} = D \begin{bmatrix} z_{i,1} \\ \ldots \\ z_{i,C} \end{bmatrix} \tag{16}$$

Here $P_i$ is the patch extraction operator at the $i^{th}$ location, $D$ is the dictionary representing all the patches. It is a big dictionary having $mn \times C$ rows (m and n being the patch dimensions) and $d \times C$ columns where d is the number of dictionary atoms used for a single image. Estimation of the image is expressed as,

$$\min_{X,D,Z} \|vec(Y) - R(I \otimes F)vec(X)\|_2^2 + \mu \left( \sum_i \left\| \begin{bmatrix} P_i x_1 \\ \ldots \\ P_i x_C \end{bmatrix} - D \begin{bmatrix} z_{i,1} \\ \ldots \\ z_{i,C} \end{bmatrix} \right\|_F^2 + \lambda \left\| \begin{bmatrix} z_{i,1} \\ \ldots \\ z_{i,C} \end{bmatrix} \right\|_1 \right) \tag{17}$$

Mathematically its solution is exactly the same as the standard dictionary learning based technique for single images. But the problem with this approach is that the size of the dictionary becomes much larger. This means that there are more parameters (dictionary elements) to estimate. Given limited training data, this is likely to overfit. Hence the result from this approach may not be the best as well.

We propose to exploit the structure of the images without increasing the size of dictionary. We express the patches of the multiple echoes in the form of a multiple measurement vector matrix.

$$X_i = [P_i x_1 | \ldots | P_i x_C] \tag{18}$$

Just as in the CS based formulation (8), we propose to learn a dictionary such that the coefficients of $X_i$ are row-sparse. The row-sparsity accounts for the common structure of the underlying echoes. In the following formulation (17), $Z_i$ is row-sparse.

$$X_i = DZ_i = D[z_{i,1} | \ldots | z_{i,C}] \tag{19}$$

In the image reconstruction stage, we recover the image by exploiting this group-sparsity among the patches from the same location. This is expressed as,

$$\min_{X,D,Z} \|vec(Y) - R(I \otimes F)vec(X)\|_2^2 + \mu \left( \sum_i \|X_i - DZ_i\|_F^2 + \lambda \|Z_i\|_{2,1} \right) \quad (20)$$

As is done in any dictionary learning based reconstruction technique, (20) is solved iteratively using alternating minimization.

$$P1: \min_X \|vec(Y) - R(I \otimes F)vec(X)\|_2^2 + \mu \sum_i \|X_i - DZ_i\|_F^2$$

$$P2: \min_D \sum_i \|X_i - DZ_i\|_F^2$$

$$P3: \min_Z \sum_i \|X_i - DZ_i\|_F^2 + \lambda \|Z_i\|_{2,1}$$

The sub-problems P1 and P2 are least squares problems. They have a closed form solution. However for large scale problems, it is inefficient to solve it in closed form. Hence we solve it using conjugate gradient. Sub-problem P3 can be separated for each patch –

$$\min_{Z_i} \|X_i - DZ_i\|_F^2 + \lambda \|Z_i\|_{2,1} \quad (21)$$

This is updated by the modified iterative soft thresholding algorithm [10].

This is a non-convex problem, there is no guarantees on global convergence. We stop the iterations depending on two criteria. The first one is a maximum number of iterations (50 for our work); the second one is the local convergence, i.e. when the change in cost function is less than some threshold.

Since the problem is non-convex, the results depend on the initial estimate. Usually papers in dictionary learning initialize it randomly – hence the results are not repeatable. Here we propose a deterministic initialization.

Estimate the image by solving $\min_X \|vec(Y) - R(I \otimes F)vec(X)\|_2^2$

From the initial estimate, compute the SVD – $[X_1 | ... | X_N] = USV^T$. Here $X_i$ is for the $i^{th}$ patch.

Initialize the dictionary from the left singular values.

Such an initialization is repeatable and always yields good results.

## 2.2. Transform Learning

The transform learning based formulation is based on the same patch-wise representation as (18). Instead of expressing the patches in terms of the dictionary as in (19), we express them in terms of the transform as follows,

$$TX_i = Z_i = [z_{i,1} | ... | z_{i,C}] \quad (22)$$

The image reconstruction is formulated in a manner similar to dictionary learning (20). The first term imposes global data consistency, the second term is for transform learning. The complete formulation is shown in (23).

$$\min_{x,T,Z} \|vec(Y) - R(I \otimes F)vec(X)\|_2^2 + \mu \left( \sum_i \|TX_i - Z_i\|_F^2 + \lambda \|Z_i\|_{2,1} + \gamma \left( \|T\|_F^2 - \log \det T \right) \right) \quad (23)$$

As in dictionary learning, we impose row-sparsity in the transform coefficients on the patches corresponding to the same location '$i$'; this accounts for the structure of the underlying multi-echo images. As before, (23) can be segregated into the following sub-problems.

$$S1: \min_x \|vec(Y) - R(I \otimes F)vec(X)\|_2^2 + \mu \sum_i \|TX_i - Z_i\|_F^2$$

$$S2: \min_T \sum_i \|TX_i - Z_i\|_F^2 + \gamma \left( \|T\|_F^2 - \log \det T \right)$$

$$S3: \min_Z \sum_i \|TX_i - Z_i\|_F^2 + \lambda \|Z_i\|_{2,1}$$

The first sub-problem S1, is a simple least squares problem with closed form solution. The second one is the standard transform update. It too has a closed form solution [23]. The third sub-problem has a closed form solution as well; unlike the case in dictionary learning where the $l_{2,1}$-minimization problem was solved iteratively, S3 has a closed form solution [10].

This concludes the derivation of our algorithm. We have imposed two stopping criteria. The first one is a maximum number of iterations – it has been fixed to 50 in all cases. The second criterion is the local convergence; iterations continue till the objective function (23) does not change significantly in subsequent iterations.

As in dictionary learning, we propose a similar deterministic initialization.

Estimate the image by solving $\min_X \|vec(Y) - R(I \otimes F)vec(X)\|_2^2$

From the initial estimate, compute the SVD – $[X_1 | ... | X_N] = USV^T$. Here $X_i$ is for the $i^{th}$ patch.

Initialize the transform from the transpose of left singular values.

## 3. Experimental Results

### 3.1. Data Acquisition

Two sets of fully sampled CPMG data were acquired for use as baselines for simulated undersampling: a) a rat lumbar spinal cord *ex vivo*, and b) rat lumbar spinal cord *in vivo*. The excised sample was also used for actual CS CPMG acquisitions.

### 3.1.1. Ex-Vivo

All animal experimental procedures were carried out in compliance with the guidelines of the Canadian Council for Animal Care and were approved by the institutional Animal Care Committee. One female Sprague-Dawley rat was obtained from a breeding facility at the University of British Columbia and acclimatized for seven days prior to the beginning of the study. Animal was deeply anaesthetized and perfused intracardially with phosphate buffered saline for 3 minutes followed by freshly hydrolysed paraformaldehyde (4%) in 0.1 M sodium phosphate buffer at pH 7.4. The 20 mm spinal cord centred at C5 level was then harvested and post-fixed in the same fixative. MRI experiments were carried out on a 7 T/30 cm bore animal MRI scanner (Bruker, Germany). Single slice multi-echo CPMG sequence (8) was used to acquire fully sampled $k$-space data from the excised spinal cord sample using a 5 turn, 13 mm inner diameter solenoid coil with $256 \times 256$ matrix size, TE/TR = 6.738/1500 ms, 32 echoes, 2.56 cm field-of-view (FOV), 1 mm slice, number of averages (NA) = 8, and the excitation pulse phase cycled between 0° and 180°. Acquisition time was 50 minutes. The high signal average (NA = 8) was used to minimize noise contribution to better assess the efficacy of our CS scheme in the simulated undersampling portion.

### 3.1.2. In-Vivo

Rectangular coil 22 x 19 mm was surgically implanted over the lumbar spine (T13/L1) of a female Sprague-Dawley rat as described previously (9). For MRI experiments, animal was anaesthetized with isofluorine (5% induction, 2% maintenance) mixed with medical air and positioned supine in a specially designed holder. Respiratory rate and body temperature were monitored using an MRI compatible monitoring system (SA Instruments, Stony Brook, NY). Heated circulating water was used to maintain the body temperature at 37ºC. Data was acquired using the same CPMG sequence but with slice thickness of 1.5 mm and in-plane resolution of 117 ☐m. The slice was positioned at T13/L1 level, and NA=6. To minimize motion artefacts the acquisition was triggered to the respiratory rate, which resulted in the total acquisition time of approximately 45 minutes.

### 3.1.3. Simulated Under-sampling / Acceleration

In this work we follow the same experimental methodology as in [2, 3 and 24]. We simulate variable density partial sampling of the K-space by randomly omitting lines in the frequency encoding direction. Two sampling patterns for 32 and 64 lines in the read-out direction are used; they correspond to sampling ratios 12.5% and 25% of the full K-space. For all the sampling patterns, a third of the total sampling lines are used to densely sample the center of the K-space. The rest of the sampling lines are spaced uniformly at random over the remaining K-space.

Both the algorithms require specifying several parameters. These parameters are tuned by the greedy L-curve method. For the first parameter the others are put to zero and the standard L-curve method is used to find the first value. For the second parameter, the first parameter is fixed at the obtained value and the third parameter is put to zero; with these setting the standard L-curve method is used to find the second parametric value. For the third parameter the first and the second values are kept fixed and the L-curve method is used to find the third parametric value.

For both the proposed techniques, the number of basis elements is kept the same size of the patches. In this case we have used patches of size 8 x 8 (standard in all reconstruction papers), therefore the number of basis is kept 64.

We have compared our result with three prior studies. The papers [2] and [3] are of similar nature, differing from each other in the synthesis [2] and analysis [3] formulations; of these [3] yields better results hence we will compare with [3]. A later work [24] improves upon the aforesaid formulations by incorporating group-sparsity. The aforesaid studies use a fixed basis. We compare with [22] which is a dictionary learning based adaptive reconstruction framework; but based only on sparsity. It does not incorporate the group-sparse structure proposed in this work. For the prior studies, the settings and configurations used to generate the best results have been used here.

## 3.2. Quantitative Results

We use the Signal-to-Noise ratio (SNR) as the metric for comparing the quantitative reconstruction results. The results are shown in Tables 1 and 2 for ex-vivo and in-vivo respectively.

In the next set of experiments, we use a different sampling mask for each echo. Here instead of using the MMV formulation, we use the proposed optimization problem (10). Therefore we compare our work with [5] – which uses the group-sparsity promoting optimization problem (5) to reconstruct the images.

Table 1. Ex-Vivo: SNR from different techniques

| Recovery Method | 32 lines | 64 lines |
| --- | --- | --- |
| Analysis row-sparsity [3] | 12.7 | 16.7 |
| Rank-deficient analysis row-sparsity [24] | 14.6 | 18.1 |
| Sparse DL formulation [22] | 16.4 | 19.2 |
| Row-sparse DL formulation | 19.2 | 23.6 |
| Row-sparse TL formulation | 20.0 | 24.2 |

Table 4. In-Vivo: SNR from different techniques

| Recovery Method | 32 lines | 64 lines |
| --- | --- | --- |
| Analysis row-sparsity [3] | 10.7 | 15.3 |
| Rank-deficient analysis row-sparsity [24] | 12.6 | 17.5 |
| Sparse DL formulation [22] | 13.7 | 18.1 |
| Row-sparse DL formulation | 16.5 | 22.3 |
| Row-sparse TL formulation | 17.1 | 23.3 |

The results are expected. By learning the basis adaptively we improve upon the standard compressed sensing based techniques by a significant margin in terms of quality; we improve by 4 to 5 dB. But what is even more worth noticing is the fact that compared to the compressed sensing based technique at 4 fold acceleration (64 lines) we produce better quality images at 8 fold acceleration (32 lines). This means that effectively we can reduce the scan time by two fold compared to prior compressed sensing methods and still yield a better quality image. The adaptive learning technique without row-sparsity yields slightly better results than CS based techniques but is considerably poorer compared to us.

## 3.3. Qualitative Results

Numerical results do not always provide information about the qualitative nature of reconstruction. Thus, we provide reconstructed and difference images for visual inspection. Owing to limitations in space, we only show the results for 32 lines (this corresponds to an acceleration factor of 8).

Echo numbers 1, 5, 9 and 13 are shown in the following figures. In Figs. 1 and 2, the ex-vivo and in-vivo reconstructed images are shown. In Figs. 3 and 4, the difference images corresponding to ex-vivo and in-vivo imaging are shown respectively. From Fig. 1 and Fig. 3 (In-vivo), we can see that the non-adaptive techniques [3, 24] yield significantly poorer results compared to the adaptive ones. Hence we do not show these poor results for ex-vivo in Fig. 2 and Fig. 4. For the ex-vivo images we only show results between the adaptive reconstruction techniques.

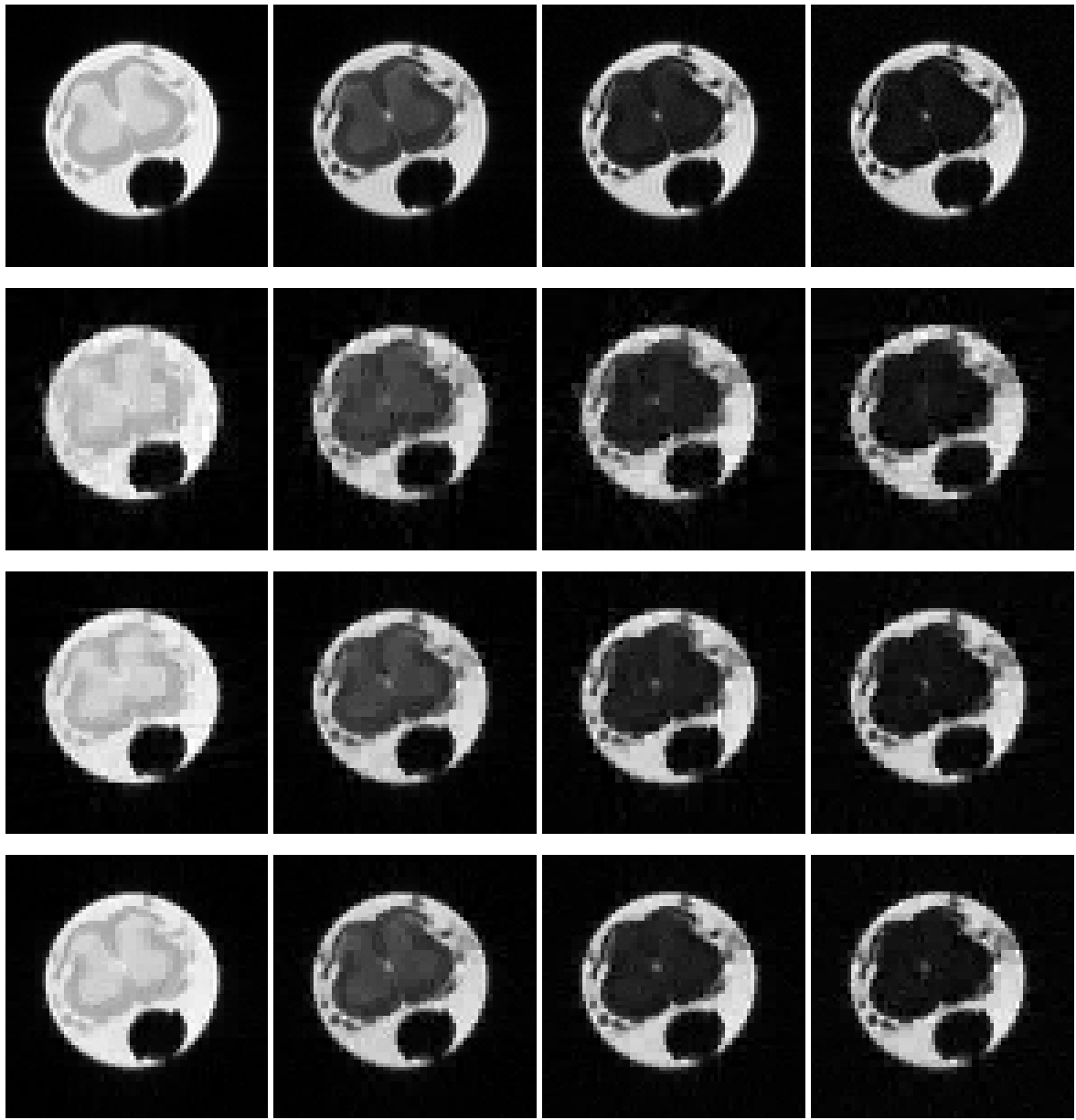

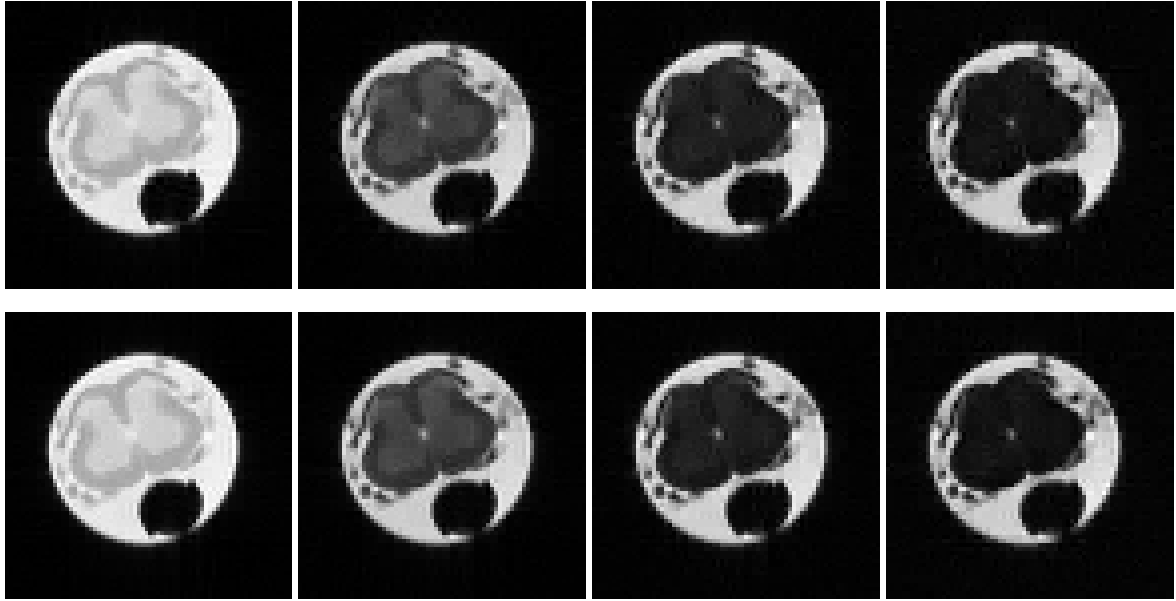

Fig. 3. Images from Ex-vivo Reconstruction. Top row –groundtruth; 2$^{nd}$ row – Analysis row-sparsity [3]; 3$^{rd}$ row – Rank deficient analysis row-sparsity [24], 4$^{th}$ row – Sparse DL formulation [22], 5$^{th}$ row – Proposed row-sparse DL formulation; 6$^{th}$ row – Proposed row-sparse TL formulation.

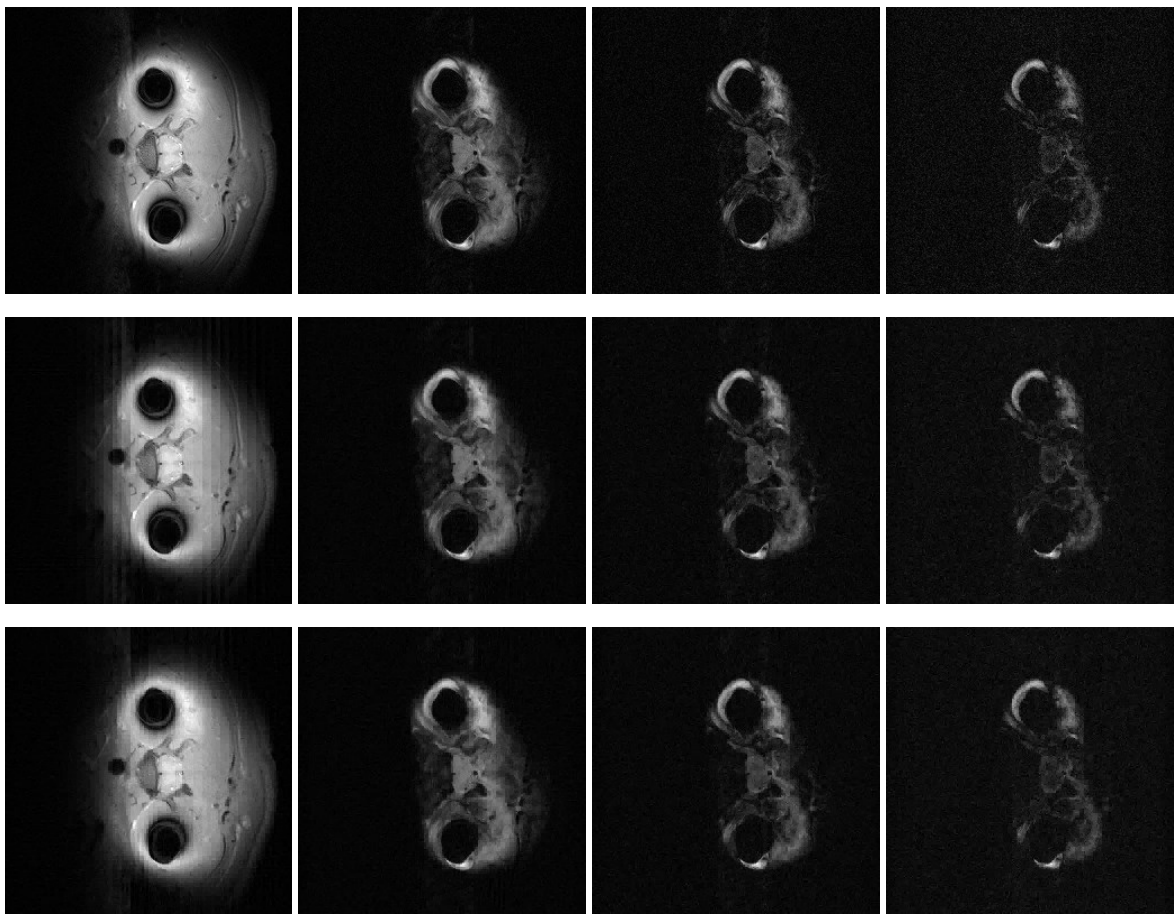

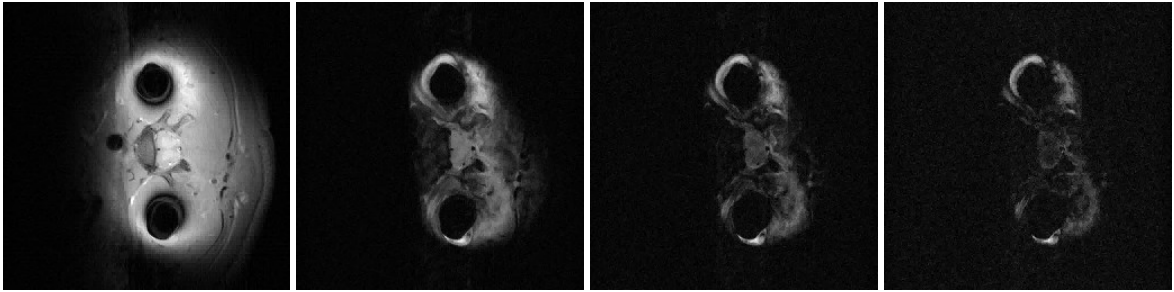

Fig. 4. Images from In-vivo Reconstruction. Top row –groundtruth; 2nd row – Sparse DL formulation [22], 3rd row – Proposed row-sparse DL formulation; 4th row – Proposed row-sparse TL formulation.

Looking closely at these figures, especially at the edges, one can see that the reconstruction quality progressively improves as one moves from top to bottom. One can easily see the poor reconstruction capability of the non-adaptive CS based techniques from rows 2 and 3 of Fig. 1. Even the adaptive technique [22] but without row-sparsity yields visible reconstruction artifacts. This can be seen from row 4 of Fig. 1 and row 2 of Fig. 2. Our proposed method yields better results visually.

The differences in the reconstruction quality are best evaluated from the difference images. The difference is formed by taking the absolute difference between the reconstructed and the ground truth images. As mentioned before, we show all the difference images for ex-vivo but only the difference images for the adaptive techniques.

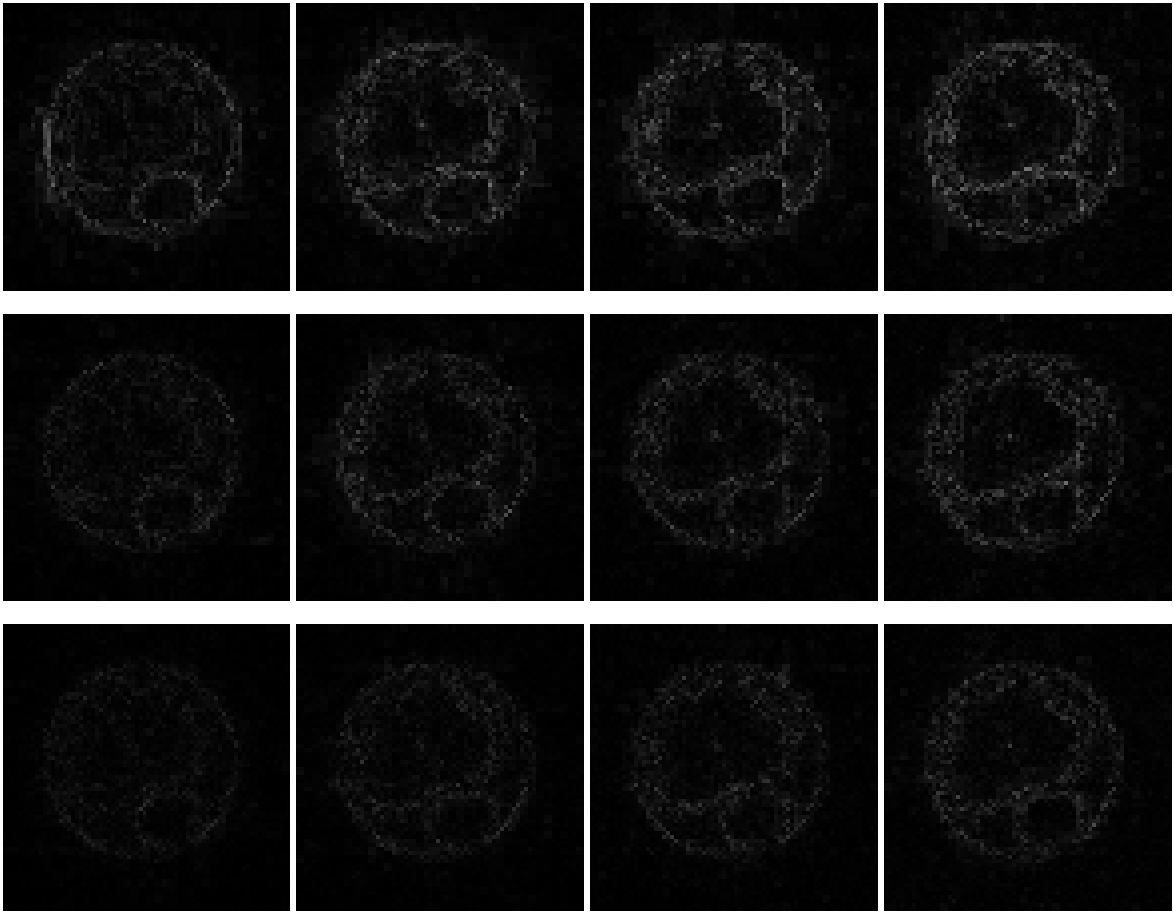

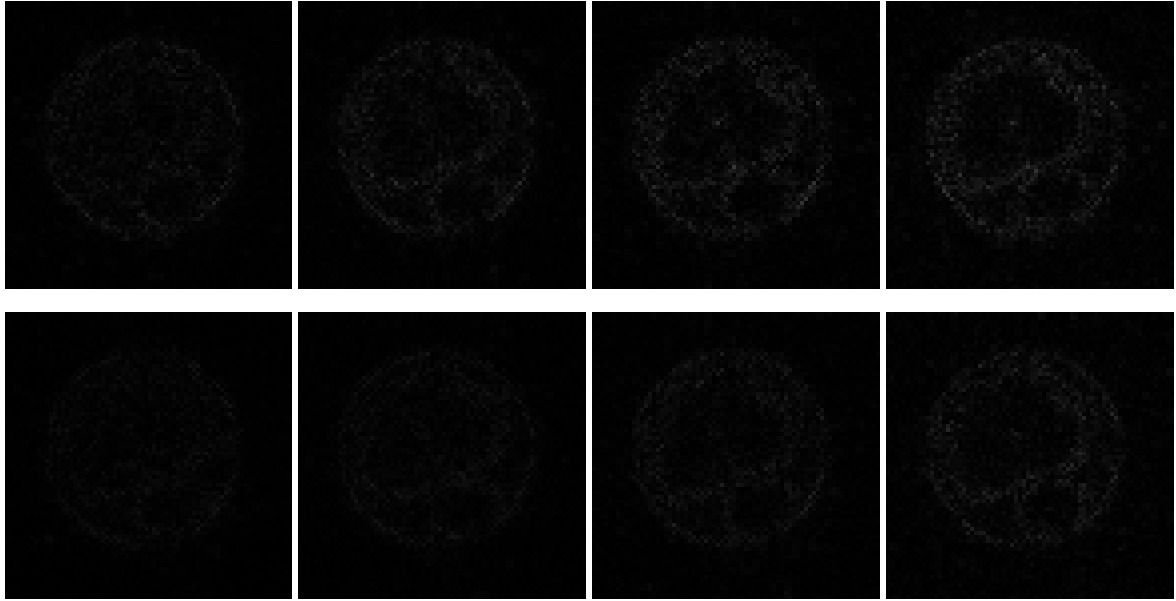

Fig. 5. Difference Images from Ex-vivo Reconstruction. Top row –Analysis row-sparsity [3]; 2nd row – Rank deficient analysis row-sparsity [24], 3rd row – Sparse DL formulation, 4th row – Proposed row-sparse DL formulation; 5th row – Proposed row-sparse TL formulation.

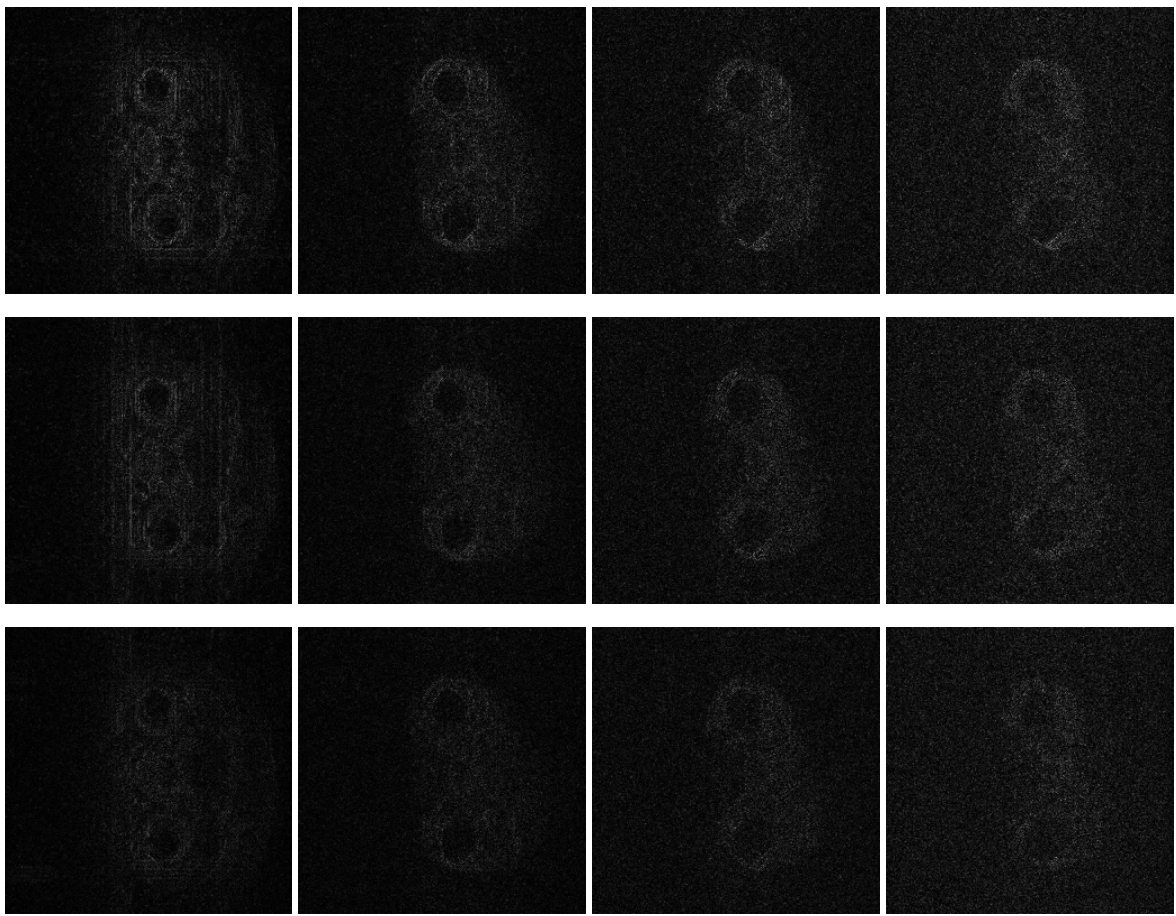

Fig. 6. Difference Images from In-vivo Reconstruction. Top row –Analysis row-sparsity [3]; 2nd row – Rank deficient analysis row-sparsity [24], 3rd row – Sparse DL formulation, 4th row – Proposed row-sparse DL formulation; 5th row – Proposed row-sparse TL formulation.

The difference in the reconstruction quality is easily discernible in the difference images. The difference images that are brighter indicate the presence of higher reconstruction error. It is clear that our proposed formulation yields considerably better results (less reconstruction artifacts) compared to the others.

## 5. Conclusion

This work addresses the problem of recovering multi-echo MRI images from their compressive K-space measurements. Prior studies used fixed basis like for recovery; here we show that by learning the basis adaptively significantly better reconstruction results can be obtained. We find that (at least for the datasets used in this work) we can produce better quality images at 8 fold acceleration compared to prior techniques at 4 fold acceleration.

There is another area in MRI that can benefit from our proposed technique. A similar problem arises in parallel MRI [25, 26] where the acquisition from different channels look similar to each other. Prior studies have used row-sparsity compressed sensing techniques for its recovery. In future, it remains to be seen if our proposed adaptive techniques can improve the recovery in parallel MRI as well.